\documentclass[a4paper]{article}

\usepackage{INTERSPEECH2020}
\usepackage{multicol}
\usepackage{graphicx}
\usepackage{multirow}
\usepackage{bibspacing}
\PassOptionsToPackage{hyphens}{url}\usepackage{hyperref}

\allowdisplaybreaks

\title{Axial Residual Networks for CycleGAN-based Voice Conversion}
\name{Jaeseong You, Gyuhyeon Nam, Dalhyun Kim, Gyeongsu Chae}
\address{
  MoneyBrain Inc., Seoul, Korea}
\email{\{jaeseongyou, ngh3053, torch, gc\}@moneybrain.ai}

\begin{document}

\maketitle
\begin{abstract}
  We propose a novel architecture and improved training objectives for non-parallel voice conversion. Our proposed CycleGAN-based model performs a shape-preserving transformation directly on a high frequency-resolution magnitude spectrogram, converting its style (i.e. speaker identity) while preserving the speech content. Throughout the entire conversion process, the model does not resort to compressed intermediate representations of any sort (e.g. mel spectrogram, low resolution spectrogram, decomposed network feature). We propose an efficient axial residual block architecture to support this expensive procedure and various modifications to the CycleGAN losses to stabilize the training process. We demonstrate via experiments that our proposed model outperforms Scyclone and shows a comparable or better performance to that of CycleGAN-VC2 even without employing a neural vocoder.
\end{abstract}
\noindent\textbf{Index Terms}: voice conversion, generative adversarial networks, CycleGAN, axial convolution

\section{Introduction}

Voice conversion (VC) is a technique to convert the speaker identity of an input speech to a different one while preserving the linguistic content. Even fairly recent VC models often necessitate a parallel dataset, a set of texts spoken twice, respectively by the source speaker and the target speaker \cite{Mohammadi01-AOV}. Such parallel corpora are hard to obtain, and the issue of aligning the tempo and the prosody between a pair of utterances further complicates the matter. Latest studies on VC therefore concern the more challenging task of learning from non-parallel utterances. Many strategies have been proposed to this end: sequence-to-sequence architectures that explicitly decompose linguistic content via automatic speech recognition (ASR) and text-to-speech (TTS) \cite{Tanaka01-A2V, Huang01-VTN}, variational autoencoder(VAE)-based methods \cite{Kameoka01-ACV, Tobing01-NPV}, the style-changing StarGAN \cite{Choi01-SUG} applied to the domain of voice \cite{Kameoka02-SNP, Kaneko02-SRC}, networks based on CycleGAN \cite{Zhu01-UIT} that cycle between two voice identities \cite{Kaneko03-PDF, Kaneko04-CNP, Kaneko05-CIC, Kaneko06-CEI}. Our proposed model is of the last category.

The most closely related work to our proposed model is Scyclone \cite{Tanaka02-SHQ}, from which it inherits a CycleGAN-based 1D-convolution-only architecture that manages a linear magnitude spectrogram directly. The authors of Scyclone claims to outperform the previous state-of-the-art CycleGAN-based voice conversion model, CycleGAN-VC2 \cite{Kaneko05-CIC}, in both similarity and naturalness. Its architecture, however, does not scale effectively with the increasing frequency resolution of spectrogram, leading to exploding gradients (confirmed in repeated primary experiments). Moreover, the model responds sensitively to a sample rate higher than 16kHz or non-parallel utterances that differ significantly in their phonetic composition; it quickly deteriorates in quality and fails to preserve speech content. Finally, the receptive field of the 1D convolutional neural network (CNN) residual block--that is commonly used across VC networks \cite{Kaneko03-PDF, Kaneko04-CNP, Kaneko05-CIC, Kaneko06-CEI}--is limited to maneuver low frequency signals with high fidelity.

Our proposed model overcomes the difficulties by introducing modifications to the architectural design of the residual block and the composition of loss functions. By carrying the spectral information without channel-wise inflation or deflation, the model evinces that even the high-resolution data can be effectively processed at once with neither the multi-level representation of audio feature in different time scales at the expense of computation and memory \cite{Kumar01-ANI, Kong01-HGA} nor the low-resolution representation at the cost of performance \cite{Tanaka02-SHQ}. This shape-preserving approach resembles that of distribution-modifying statistical models \cite{WaveGlow, WaveGrad}, opening up possible extensions to flow or diffusion, but such discussion is outside the scope of this paper. Our major contributions here are four-fold:
1) proposing an efficient axial residual block architecture with an extended receptive field in the temporal axis
2) utilizing a high frequency-resolution spectrogram as input without resorting to a downscaled spectrogram
3) enabling the direct spectrogram-to-spectrogram conversion that preserves the data shape end-to-end both time- and channel-wise
4) introducing various modifications to the CycleGAN loss that stabilize the training process and improve the fidelity of the resulting voice.

\section{Model Description}

\subsection{Axial Residual Block}

\begin{figure}[ht]
  \centering
  \includegraphics[scale=0.4]{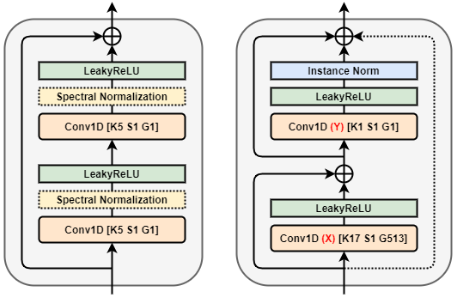}
%   \vspace*{-5mm}
  \caption{The conventional 1D-convolution residual block (left) and the proposed axial residual block (right). The dotted lines indicate optional components, and k, s, g denote kernel, stride, and group respectively. The axial block in the figure is equipped with a depth-wise convolution to process x-axis information, which can be further streamlined at a minimal compromise in performance if replaced with lightweight convolution. The y-axis convolution, while illustrated to have the kernel size of 1, can be extended to a larger kernel for optimal performance.}
  \label{fig:axial_residual_block}
\end{figure}

The conventional residual block architecture (employed in both CycleGAN-VC and Scyclone) is not effective in the challenging task of transforming one form of high frequency-resolution spectrogram to another without intermediate compressed representation, due to its limited receptive field. To overcome this difficulty, our proposed axial residual block is composed of the temporal axis convolution layer and the frequency axis convolution layer. Figure~\ref{fig:axial_residual_block} illustrates the details. The architectural design bears high similarity to the axial attention \cite{Ho01-AAM, huang2019ccnet} in that the receptive field is efficiently widened in two-step axial operation by capturing information along one of the two axes at a time. The major difference from such self-attention variants, however, is that the kernel is not dynamically computed every time frame, which allows to process an input of variable length unlike in a Transformer block where the input length should be fixed in advance. It also resembles spatially separable convolution where the expressivity of one convolution layer is approximated in two steps for computational efficiency \cite{Mamalet01-SCF}.

The authors of \cite{Dosovitskiy01-AII} observe that there is a qualitative difference between the receptive field widened over layers and the one that is wide from the beginning, emphasizing the advantage of the latter. In addition, Kong et al. underline that phonemic relations in speech often exceed over 100ms \cite{Kong01-HGA}. Hence, an ideal VC model should be able to process signal information that lie over the 100ms window from the very first layer, instead of reaching the scope in successive layers. For the temporal axis (i.e. x-axis) convolution layer, we therefore employ depth-wise convolution so that each kernel learns to process the changes in one frequency band independently of others. This computational efficiency allows an extremely large kernel size. Our choice is 17, which translates to 230ms (or 4.35Hz given the 22.05kHz sample rate). Every block thus can capture long-term correspondences (low-frequency signals) that occur across over a couple of thousand samples, and we view this capability to be critical for the improved performance.

A viable alternative to this depth-wise convolution for transformation in the time domain is the lightweight convolution \cite{Wu01-PLA}. While it is highly similar to the depth-wise 1D convolution where each kernel is learned channel-wise, the lightweight convolution further improves efficiency by sharing the same kernel over a specified number of channels. First proposed in the domain of natural language processing, it has been successfully adopted in processing audio information \cite{Elias01-PTN, Baevski01-W2V}; when applied to the domain of audio, the inductive bias of the 1D lightweight convolution is that information in nearby frequency bands can be effectively processed with a shared operation, which well approximates the workings of human auditory cognition.

The frequency-axis (i.e. y-axis) convolution layer is a standard 1x1 1D convolution; the only particularity is that the number of kernels is the same as the number of input channels. This is in fact equivalent to a position-wise feedforward layer, resulting in kernels that focus on processing the information at one specific time-point only. The inductive bias here is that changing one voice to another can be successfully realized by applying one nonlinear transformation communally across all the spectral frames. To increase the capacity and further widen the overall temporal receptive field, we enlarge the kernel size to 3 for our experiments. The vertical strap thus obtains more expressivity regarding temporal locality.

Residual connections can be made either once between the input and the output of the block, which we choose for our experiments, or twice between the input and the outputs of the two convolution layers. With the latter, the axial residual block attains closer connections to the original Transformer block \cite{Vaswani01-AIA}. The only difference lies in that it employs either the depth-wise or lightweight convolution layer instead of the self-attention layer. In fact, replacing the self-attention layer of the Transformer block with dynamic lightweight convolution has been proposed in \cite{Yun01-ATU}, showing improvements when applied to early layers. One can interpret our proposed model as its static variant where the kernel is static across the temporal dimension and the replacement is applied to the entire stack instead of only to early layers.

\subsection{Improved Objective}

While CycleGAN-based VC models often employ a global discriminator that outputs one prediction averaged over the entire utterance (presumably for stable training), it tends to lead to various artifacts in the synthesized output: notably, outbursts of phonemic fragments and momentary distortion of content. We hypothesize that this is because the generator brings in short non-sensible fragments from the source audio to take advantage of the discriminator, which can yield only global predictions. To reduce these artifacts, we employ a patchGAN style loss instead. The effectiveness of the patchGAN adversarial loss is consistent with the findings in speech-related GAN models \cite{Kumar01-ANI, Kong01-HGA, Kaneko05-CIC}, and we thus take the patchGAN form to the extreme with regard to the temporal dimension, outputting a discriminative prediction per spectrogram frame.

Adversarial loss, while being the key component in transforming the style, tends to induce serious instability during training. CycleGAN-based VC models thus tend to use a modified GAN loss such as least square loss \cite{Mao01-LSG} or hinge loss \cite{Lim01-GGA} to curb down the high volatility. In adjoinment with the PatchGAN design \cite{Isola01-IIT} extended to the maximal resolution in the time domain, however, our preliminary experiments indicate that using a plain binary cross entropy objective is more effective than the aforementioned variants. The adversarial loss in our proposed model is therefore defined as follows:
\begin{align}
L_{adv}(&G_{X \to Y}, D_{Y}) =\nonumber \\
&\mathbb{E}_{y \sim P_{Y}(y)}[\log D_{Y}(y)] \\ 
&+ \mathbb{E}_{x \sim P_{X}(x)}[\log (1 - D_{Y}(G_{Y \to X})] \nonumber
\label{eq1}
\end{align}

The cycle-consistency loss penalizes the case where the result converted twice severely deviates from the original utterance (i.e. after the style has changed from the source to the target, and back to the source once more). The original consistency loss is defined as follows:
\begin{align}
L_{cyc}(&G_{X \to Y}, G_{Y \to X}) =\nonumber \\
&\mathbb{E}_{x \sim P_{X}(x)}[||G_{Y \to X}(G_{X \to Y}(x)) - x||_{1}]\\
&+ \mathbb{E}_{y \sim P_{Y}(y)}[||G_{X \to Y}(G_{Y \to X}(y)) - y||_{1}] \nonumber
\label{eq2}
\end{align}

However, the content reconstruction often breaks when the two training corpora differ much in terms of phonetic composition (inter-language in the extreme case) and style (inter-sex in the extreme case). We therefore augment the loss with auxiliary feature matching. In other words, we add a stronger constraint that all the pairs of the residual block outputs of the corresponding discriminator should be similar for the cycled output and the original utterance:
\begin{align} 
L&_{cyc}(G_{X \to Y},G_{Y \to X}) =\nonumber \\
&\mathbb{E}_{x \sim P_{X}(x)}[||G_{Y \to X}(G_{X \to Y}(x)) - x||_{1}] \nonumber \\
&+ \mathbb{E}_{y \sim P_{Y}(y)}[||G_{X \to Y}(G_{Y \to X}(y)) - y||_{1}] \\
&+ \mathbb{E}_{x \sim P_{X}(x)}[||D_{X}^{f}(G_{Y \to X}(G_{X \to Y}(x))) - D_{X}^{f}(x)||_{1}] \nonumber \\
&+ \mathbb{E}_{y \sim P_{Y}(y)}[||D_{Y}^{f}(G_{X \to Y}(G_{Y \to X}(y))) - D_{Y}^{f}(y)||_{1}] \nonumber
\label{eq3}
\end{align}

The identity loss, first proposed in \cite{Taigman01-TPM}, regularizes the generator to approximate an identity mapping when real samples of the target domain are fed. The expected effects are that the model conservatively responds to an unknown distribution and better preserves phonetic contents in general. The original identity loss is defined as follows:
\begin{align}
L_{id}(&G_{X \to Y},G_{Y \to X}) = \nonumber \\
&+ \mathbb{E}_{y \sim P_{Y}(y)}[||G_{X \to Y}(y) - y||_{1}] \\
&+ \mathbb{E}_{x \sim P_{X}(x)}[||G_{Y \to X}(x) - x||_{1}] \nonumber
\label{eq4}
\end{align}

We extend the identity loss by applying the identity-mapping not only for real samples, but also for the transformed utterances. In this manner, the generator is exposed to additional types of unknown distributions: 
\begin{align}
L&_{id}(G_{X \to Y},G_{Y \to X}) = \nonumber \\
&\mathbb{E}_{y \sim P_{Y}(y)}[||G_{X \to Y}(y) - y||_{1}] \nonumber \\
&+ \mathbb{E}_{x \sim P_{X}(x)}[||G_{Y \to X}(x) - x||_{1}] + \\
&+ \mathbb{E}_{x \sim P_{X}(x)}[||G_{X \to Y}(G_{X \to Y}(x)) - G_{X \to Y}(x)||_{1}] \nonumber \\
&+ \mathbb{E}_{x \sim P_{X}(x)}[||G_{Y \to X}(G_{Y \to X}(y)) - G_{Y \to X}(y)||_{1}] \nonumber
\label{eq5}
\end{align}

The total loss therefore consists of the three aforementioned components: the adversarial loss, the cycle-consistency loss, and the identity loss. The three components are linearly scaled with their corresponding lambda values to control their relative contribution to the total loss: 
\begin{equation}
L_{total} = \lambda_{adv} * L_{adv} + \lambda_{cyc} * L_{cyc} + \lambda_{id} * L_{id} 
\label{eq6}
\end{equation}

\subsection{Generator and Discriminator}

The generator architecture is fully 1D-CNN. The prenet convolution layer, a 1x1 convolution, mixes the channel information in the same manner for every time point. The intermediate stack consists of 7 axial residual blocks. Within each block, in between the temporal convolution and the frequency convolution, non-linearity is introduced with leaky ReLU activation with the slope of 0.01. The postnet convolution once again mixes the channel information followed by the final relu activation, which facilitates the generator to synthesize a linear magnitude spectrogram that should take only positive values.

\begin{figure}
  \centering
  \includegraphics[scale=0.4]{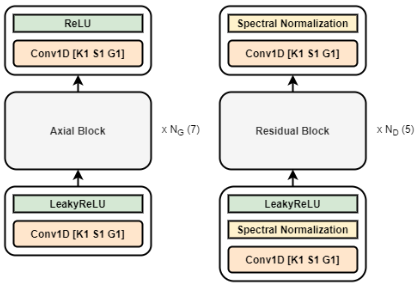}
  \setlength{\belowcaptionskip}{-15pt}
%   \vspace*{-5mm}
  \caption{Schematic architecture of generator (left) and discriminator (right). The number of blocks and the detailed configuration of each block can be further fine-tuned for optimal performance.}
  \label{fig:generator&discriminator}
\end{figure}

The architectural design of the discriminator is mostly equivalent to that of Scyclone \cite{Tanaka02-SHQ}. Gaussian noise with the standard deviation of 0.01 is added to the input to stabilize the training process \cite{Arjovsky01-TPM}. Additionally, spectral normalization is applied to the weights of all the convolution layers \cite{Miyato01-SNF}, to prevent the discriminator from learning too quickly by capping the magnitude of possible change: 
\begin{equation} 
W_{SN} = \frac{W}{\sigma(W)}, \sigma(W) = max_{h:h \neq 0} \frac{||Wh||_{2}}{||h||_{2}}
\label{eq7}
\end{equation}
The subsequent structure following the prenet layer consists of 5 original residual blocks. Each of them introduces non-linearity with leaky ReLU with the larger slope of 0.2. The logit values are then output from the postnet convolution. Figure~\ref{fig:generator&discriminator} shows the overall architecture.

\section{Experiments}

\subsection{Experimental Setup}
We compare our proposed model to CycleGAN-VC2 \cite{Kaneko05-CIC} and Scyclone \cite{Tanaka02-SHQ} for English and Korean voice conversion. For the former, we use the multi-speaker VCTK dataset \cite{VCTK}. Mic1 recordings of two females (p299 and p301) and two males (p311 and p360) are selected and downsampled to 22.05kHz (for our proposed model) or to 16kHz (for CycleGAN-VC2 and Scyclone). The 24 parallel utterances shared by the four speakers are reserved for evaluation, and the rest are utilized for training. We normalize the audio, trim its beginning silence, and pad zeros at the end if needed to meet the following data shape. The 128-frame 513-channel linear magnitude spectrogram is computed using the window length of 1024 and the hop size of 256 samples with Hann window. 

For the latter Korean voice conversion, we use Korean Single-Speaker dataset (KSS) \cite{KSS} and our proprietary Korean female dataset (JEY). After removing excessively short samples, we use 11,000 utterances per identity for training, reserving 584 and 372 samples for evaluation, which are sampled randomly since the two datasets share no parallel utterances. The remaining audio preprocessing procedures are identical to those of the former experiment.

The two experiments share the same training process. Two pairs of discriminator and generator are initiated. One generator converts voice from identity X to identity Y, and the other does the opposite. The discriminator paired with the former tells whether a voice sample of identity Y is real or converted from identity X, and the one paired with the latter does the same for the identity X. For both the generators and the discriminators, we use the Adam optimizer \cite{Kingma01-AAM} with $\alpha = 2.0 \times 10^{-4}$, $\beta_{1} = 0.5$, and $\beta_{2} = 0.999$. The learning rates are annealed by a factor of $.1$ every 50 epochs. Each model is trained for 200 epochs with the batch size of 16. The scaling factors of the total loss  $\lambda_{adv}$, $\lambda_{cyc}$, and $\lambda_{id}$ are set to $1.$, $10.$, and $1.$ respectively. For Scyclone and CycleGAN-VC2, we follow the same training setting as proposed in the original papers.

Finally, when converting a resulting spectrogram to a waveform, one can expect a substantial performance gain from using a neural vocoder (e.g. WaveGlow\cite{WaveGlow}, WaveGrad\cite{WaveGrad}). We, however, compute the final audio with the griffin-lim algorithm \cite{GriffinLim} with 32 iterations. This is to compare the models on an equal footing. CycleGAN-VC2, however, is inherently dependent on the use of the WORLD vocoder \cite{Morise01-WAV}, so we make an exception for the model.

\begin{figure*}[h]
  \begin{multicols}{3}
    \includegraphics[scale=0.365]{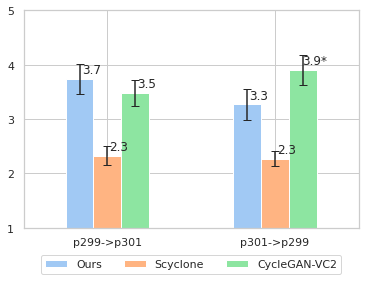} 
    \includegraphics[scale=0.365]{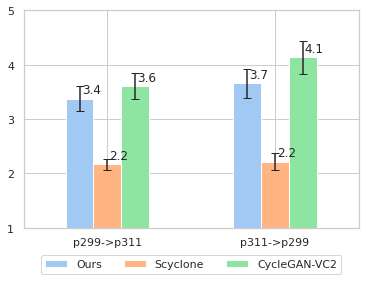} 
    \includegraphics[scale=0.365]{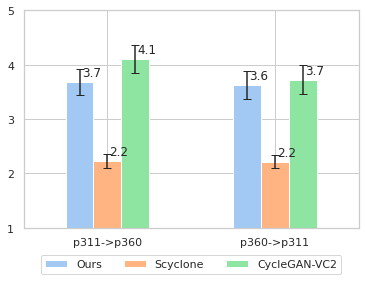}
  \end{multicols}
  \setlength{\belowcaptionskip}{-15pt}
  \vspace*{-5mm}
  \caption{MOS test on three pairs of the four selected VCTK speakers.}
  \label{fig:eng_MOS}
\end{figure*}

\subsection{Results}
To compare the structural similarity of the converted outcome and the ground truth, we use mel spectral distortion (MSD) metrics. While a converted utterance does not perfectly align with its true correspondence, we apply dynamic time warping (DTW) to pair positions where the two sequences are most similar. For the computation of the metrics, we use 40-channel log-mel spectrogram based on a 46.44ms Hann window and a hop size of 11.61ms. The mel filters span from 0 to 8,000 Hz. In the case of the Korean voice conversion experiment where the DTW sequence paring is impossible due to the lack of parallel utterances, we compare the MSD of each of the converted utterances to the ground truth MSD averaged per identity.

Word error rate (WER) measures how well each model preserves the linguistic content during the conversion process, and it is computed based on the comparison of the ASR result of the converted utterance to the ground truth text. We select commercial ASR models based on the overall stability for each experiment, \cite{Azure} for VCTK and \cite{Clova} for Korean, respectively.

In addition to this objective evaluation, we conduct human listening tests for subjective evaluation of the converted speech. Each of the 10 subjects is asked to provide his or her mean opinion score (MOS)\footnote{Audio samples are at \url{https://moneybrain-research.github.io/axial-residual-networks-vc}.} (5: very similar to 1: very dissimilar), evaluating how similar the converted speech is to the real utterance of the target identity. A rater evaluates a pair of a true utterance and a converted speech, repeating for 24 utterances per voice identity for the VCTK experiment. In the Korean experiment, we use 10 randomly sampled instances per identity. 

\begin{table}[th]
  \caption{Average MSD and WER results on VCTK voice conversion. The results are organized by the conversion from the former person ID to the latter person ID.}
  \label{tab:eng_MSD}
  \centering
  \begin{tabular}{cllll}
    \toprule  
                        &     & Scyclone              & VC2                     & ours                    \\
    \midrule
    \multirow{1}{*}{299-301}  & MSD & 2.55 $\pm$ .18  & \textbf{2.18} $\pm$ .17 & 2.38 $\pm$ .22          \\
                              & WER & 0.94 $\pm$ .04  & \textbf{0.06} $\pm$ .03 & 0.07 $\pm$ .03          \\
    \multirow{1}{*}{301-299}  & MSD & 2.26 $\pm$ .18  & 2.22 $\pm$ .22          & \textbf{2.18} $\pm$ .22 \\
                              & WER & 0.95 $\pm$ .04  & \textbf{0.07} $\pm$ .04 & 0.10 $\pm$ .08          \\ 
    \midrule
    \multirow{1}{*}{311-360}  & MSD & 1.99 $\pm$ .13  & 1.83 $\pm$ .16          & \textbf{1.81} $\pm$ .16 \\
                              & WER & 0.97 $\pm$ .02  & \textbf{0.11} $\pm$ .05 & 0.15 $\pm$ .07          \\ 
    \multirow{1}{*}{360-311}  & MSD & 2.29 $\pm$ .18  & \textbf{2.04} $\pm$ .15 & 2.06 $\pm$ .15          \\
                              & WER & 0.99 $\pm$ .01  & 0.10 $\pm$ .04          & \textbf{0.08} $\pm$ .04 \\
    \midrule
    \multirow{1}{*}{299-311}  & MSD & 2.32 $\pm$ .16  & 2.08 $\pm$ .18          & \textbf{2.06} $\pm$ .15 \\
                              & WER & 0.98 $\pm$ .01  & \textbf{0.10} $\pm$ .05* & 0.33 $\pm$ .08         \\ 
    \multirow{1}{*}{311-299}  & MSD & 2.71 $\pm$ .19  & \textbf{2.32} $\pm$ .27 & 2.58 $\pm$ .18          \\
                              & WER & 0.98 $\pm$ .02  & \textbf{0.12} $\pm$ .06* & 0.30 $\pm$ .10         \\
    \bottomrule
  \end{tabular}
\end{table}

\begin{figure}[t]
  \centering
  \includegraphics[scale=0.365]{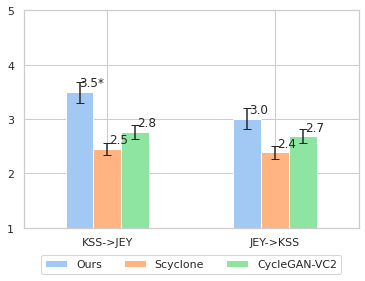}
  \setlength{\belowcaptionskip}{-15pt}
%   \vspace*{-5mm}
  \caption{MOS test on KSS-JEY voice conversion and vice versa.}
  \label{fig:kor_MOS}
\end{figure}

In terms of changing the voice identity, Table~\ref{tab:eng_MSD} and Figure \ref{fig:eng_MOS} show that our proposed model outperforms Scyclone and achieves a comparable performance to that of CyclegGAN-VC2 in the VCTK experiment. An asterisk denotes a statistically significant top result. Table~\ref{tab:kor_MSD} and Figure~\ref{fig:kor_MOS} summarize the voice conversion experiment on Korean, in which our proposed model outperforms both Scyclone and CyclegGAN-VC2 in changing styles even without a neural vocoder. We hypothesize that, since our proposed model handles high-resolution data without compression, it is more data-greedy to train to a sufficient level. VCTK, where only less than 400 utterances are available per identity, is thus not prone to fully realizing the capability of our proposed architecture. While CycleGAN-VC2 consistently does well on WER, we suspect this is because the model tends to introduce changes to style conservatively.

\begin{table}[th]
  \caption{Average MSD and WER results on KSS-JEY voice conversion and vice versa.}
  \label{tab:kor_MSD}
  \centering
  \begin{tabular}{cllll}
    \toprule  
                        &       & Scyclone                  & VC2                     & Ours                  \\
    \midrule
    \multirow{1}{*}{KSS-JEY}  & MSD   & 1.63 $\pm$ .50      & 1.61 $\pm$ .24          & \textbf{1.50} $\pm$ .45 \\
                              & WER   & 0.85 $\pm$ .25      & \textbf{0.04} $\pm$ .06* & 0.30 $\pm$ .12          \\
    \midrule
    \multirow{1}{*}{JEY-KSS}  & MSD   & 1.73 $\pm$ .42      & \textbf{1.32} $\pm$ .29 & 1.39 $\pm$ .41          \\
                              & WER   & 0.75 $\pm$ .33      & \textbf{0.21} $\pm$ .20 & 0.36 $\pm$ .19          \\
    \bottomrule
  \end{tabular}
\end{table}

\section{Conclusions}

We propose a fully 1D CNN network to process a high frequency-resolution spectrogram directly with its shape unchanged. In comparison to the conventional residual block employed across CycleGAN-based VC models, our proposed axial residual block extends the temporal receptive field while lowering memory and computation requirements. We address meaningful similarities and connections to various recent architectural solutions such as axial attention and Transformer block. The effectiveness of the proposed design is demonstrated through various experiments; our model outperforms Scyclone on all the evaluation metrics. It shows comparable or better performance to that of CycleGAN-VC2 even without a neural vocoder, especially in terms of speaker similarity with larger Korean datasets.

\bibliographystyle{IEEEtran}

\bibliography{mybib}

% \begin{thebibliography}{9}
% \bibitem[1]{Davis80-COP}
%   S.\ B.\ Davis and P.\ Mermelstein,
%   ``Comparison of parametric representation for monosyllabic word recognition in continuously spoken sentences,''
%   \textit{IEEE Transactions on Acoustics, Speech and Signal Processing}, vol.~28, no.~4, pp.~357--366, 1980.
% \bibitem[2]{Rabiner89-ATO}
%   L.\ R.\ Rabiner,
%   ``A tutorial on hidden Markov models and selected applications in speech recognition,''
%   \textit{Proceedings of the IEEE}, vol.~77, no.~2, pp.~257-286, 1989.
% \bibitem[3]{Hastie09-TEO}
%   T.\ Hastie, R.\ Tibshirani, and J.\ Friedman,
%   \textit{The Elements of Statistical Learning -- Data Mining, Inference, and Prediction}.
%   New York: Springer, 2009.
% \bibitem[4]{YourName17-XXX}
%   F.\ Lastname1, F.\ Lastname2, and F.\ Lastname3,
%   ``Title of your INTERSPEECH 2020 publication,''
%   in \textit{Interspeech 2020 -- 20\textsuperscript{th} Annual Conference of the International Speech Communication Association, September 15-19, Graz, Austria, Proceedings, Proceedings}, 2020, pp.~100--104.
% \end{thebibliography}

\end{document}